\documentclass[%
 reprint,
superscriptaddress,
 amsmath,amssymb,
 aps,
prb,
floatfix,
]{revtex4-1}

\usepackage{graphicx}
\usepackage{diagbox}
\usepackage{tikz}
\usepackage{xspace}
\usepackage{dcolumn}
\usepackage{bm}
\usepackage[colorlinks=true,allcolors=blue]{hyperref}
\usepackage{array,booktabs,tabularx}
\usepackage[caption=false]{subfig}
\usepackage[USenglish]{babel}
\usepackage{braket}
\usepackage{xcolor}
\usepackage[normalem]{ulem}
\usepackage{notes2bib}
\usepackage{multirow}
\usepackage{array}
\usepackage{alphabeta}
\usepackage[titletoc]{appendix}
\usepackage[utf8]{inputenc}
\usepackage[T1]{fontenc}
\usepackage{float}

\usetikzlibrary{positioning}
\newcolumntype{P}[1]{>{\centering\arraybackslash}p{#1}}
\newcolumntype{M}[1]{>{\centering\arraybackslash}m{#1}}

\begin{document}

\title{From superconductivity to non-superconductivity in LiPdH: a first principle approach}

\author{Zahra Alizadeh}
\affiliation{Superconductivity Research Laboratory (SRL), Department of Physics, University of Tehran, North Kargar Av., P.O. Box 14395-547, Tehran, Iran}

\author{Yue-Wen Fang}
\affiliation{Centro de F{\'i}sica de Materiales (CFM-MPC), CSIC-UPV/EHU, Manuel de Lardizabal Pasealekua 5, 20018 Donostia/San Sebasti{\'a}n, Spain}

\author{Ion Errea}
\email{ion.errea@ehu.eus}
\affiliation{Fisika Aplikatua Saila, Gipuzkoako Ingeniaritza Eskola, University of the Basque Country (UPV/EHU), Europa Plaza 1, 20018 Donostia/San Sebasti{\'a}n, Spain }
\affiliation{Centro de F{\'i}sica de Materiales (CFM-MPC), CSIC-UPV/EHU, Manuel de Lardizabal Pasealekua 5, 20018 Donostia/San Sebasti{\'a}n, Spain}
\affiliation{Donostia International Physics Center (DIPC), Manuel de Lardizabal Pasealekua 4, 20018 Donostia/San Sebasti{\'a}n, Spain
}

\author{M.R. Mohammadizadeh}
\email{zadeh@ut.ac.ir}
\affiliation{Superconductivity Research Laboratory (SRL), Department of Physics, University of Tehran, North Kargar Av., P.O. Box 14395-547, Tehran, Iran}

\date{\today}
\begin{abstract}
The layered structure of LiPdH was theoretically suggested to be a superconductor as a result of its larger electron-phonon coupling constant compared to that of PdH. However, the experimental results reported contrary findings, with no trace of superconductivity. We study the electronic, vibrational, and superconducting properties of the ambient pressure tetragonal phase of LiPdH ($P4/mmm$) within first principles density functional theory methods, both in the harmonic and anharmonic approximations for the lattice dynamics, and conclude that it does not show any superconducting behavior. High-pressure crystal structure prediction calculations indicate that no structural transition is expected to occur under pressure up to 100 GPa in LiPdH. Our theoretical calculations demonstrate that increasing pressure reduces the density of states at the Fermi surface and consequently weakens electron-phonon interactions, leading to a further suppression of the superconducting critical temperature.
\end{abstract}

\maketitle

\section{Introduction}
Superconductivity, a fundamental physical phenomenon, has intrigued researchers since its discovery in 1911 by Kammerling Onnes \cite{ROGALLA2008}. After extensive efforts to understand superconductivity across various material families, such as cuprate \cite{Park1995, mohammadizadeh2003pr, aghabagheri2018high, rasti2023superconducting}, iron-based \cite{Aswathy2010}, interface \cite{richter2013interface}, and organic superconductors \cite{jerome1991physics}, attention turned to the Ashcroft's prediction that hydrogen has the potential for room-temperature superconductivity due to the coupling of its high-frequency vibrations with electrons \cite{ashcroft1968metallic}. However, experimental challenges persist due to the requirement of high pressures above 500 GPa for hydrogen metallization \cite{loubeyre2020synchrotron,monacelli_quantum_2023}. The discovery of superconductivity in hydrogen sulfide \cite{drozdov2015conventional} marked a significant milestone in this pursuit. Chemical pressure emerged as a viable approach to achieve the metallic state of hydrogen. Consequently, a vast number of theoretical and experimental studies began to explore the superconductivity of binary hydrides. Notable studies include La-H \cite{somayazulu2019evidence, drozdov2019superconductivity}, Th-H \cite{semenok2019synthesis}, Y-H \cite{liu2017potential}, Ce-H \cite{salke2019synthesis}, and Ca-H \cite{wang2012superconductive}, which have all been confirmed experimentally. 

Following the notable outcomes observed in binary hydrides, researchers have shifted their focus to ternary compounds with the hope of reducing the pressure required to achieve superconductivity. However, the freedom to choose elements for ternaries introduces a vast array of possibilities, far exceeding 4,000. Considering all potential stoichiometries under various pressures expands this number even further, demanding a significant amount of time for a thorough examination. This diversity leads to a wide variety of structures and properties, which can exhibit interesting behaviors. Therefore, a good starting point for exploring the ternaries is to use high-temperature binary compounds as the parent structures for incorporating the third element. In essence, introducing a third element represents another step towards achieving superconductivity under ambient conditions. Some studies have suggested that substantially lower pressures are required to stabilize ternary hydrides compared to binaries. For instance, compounds like LaBeH$_8$ \cite{song2023stoichiometric} and similar ones, such as La-X-H (X=Ce, Y, Al, B,…) \cite{kostrzewa2020lah10, di2022first, chen2024high, belli2022impact}, stabilize at lower pressures than their parent La-H structure. Another example is the Ac-H system, where the inclusion of Be reduces the required pressure to keep the system dynamically stable down to 10 GPa while keeping a T$_c$ of 181 K \cite{gao2024prediction}. It appears that reaching superconductivity at lower pressures may indeed become a reality in ternary hydrides. In fact, recent first-principles calculations have predicted several new hydrides with high critical temperatures at ambient pressure that become thermodynamically metastable at moderate pressures in the order of few tens of GPa~\cite{Tiago-AFM-hydrides2024}, among which Mg$_2$IrH$_6$~\cite{sanna2023-Mg2IrH6,dolui2023-Mg2IrH6-PRL2024} and RbPH$_3$~\cite{RbPH3-arxiv2024-Dangic} seem particularly promising.


Long before the extraordinary results in hydrides under pressure, superconductivity had already been discovered at ambient pressure in hydrogen-rich compounds but with moderate critical temperatures. The first discoveries arrived in the seventies with Th$_4$H$_{15}$ and PdH \cite{satterthwaite1970superconductivity, skoskiewicz1974isotope}. 
Studying PdH$_x$ compounds at various values of hydrogen content $x$ has proven experimentally that PdH$_x$ has an observable transition to a superconducting phase when $x$ is above 0.80, and the critical temperature rises with $x$  \cite{skoskiewicz1972superconductivity, schirber1984superconductivity}. This illustrates an instance of hydrogen-enhanced superconductivity. Furthermore, Stritzker's study on ternary  Pd-M-H hydrides, incorporating noble metals such as M= Cu, Ag, and Au, unveiled that the inclusion of a third element could enhance T$_c$ \cite{stritzker1974high}. Given the renewed interest in superconducting hydrides in recent years, particularly in low-pressure ternary systems, these ambient pressure structures have regained interest. The large pressure difference between these systems and high-T$_c$ hydrides suggests a different role of hydrogen and offers new avenues for engineering novel superconductors~\cite{belli_strong_2021}. As an example, a recent \emph{ab initio} study on the ternary PdCuH$_x$ compound, related to the well-known superconducting PdH \cite{Vocaturo2022}, finds that low hydrogen content results in no superconductivity above 1 K, while high hydrogenation leads to a superconducting transition temperature of 34 K. However, the problem with the \emph{ab initio} calculations on these ambient pressure hydrides, where hydrogens sit at the interstitial sites of a compact metal lattice, is that hydrogen lattice vibrations are very anharmonic, questioning all the results obtained on superconducting~\cite{Errea2013} and even thermodynamic~\cite{meninno2023abinitio} properties.


Among ambient pressure hydrides, LiPdH not only serves as a promising candidate for investigating vibrational properties due to its hydride nature, but also its layered structural similarity to cuprates garnered significant attention. According to electronic structure results based on the local density approximation (LDA) and the McMillan-Hopfield equation, LiPdH should be an ionic superconductor comparable in T$_c$ to PdH \cite{singh1990possiblity}. In contrast, the first experimental study found almost stoichiometric LiPdH, which crystallizes in a tetragonal structure, to be non-superconducting above 4 K~\cite{noreus1990absence}.
 A second experimental study aimed at verifying the theoretical predictions managed to synthesize LiPdH under high pressure, but again no superconductivity was observed down to 2 K~\cite{Liu2017}.


In this work, we solve the discrepancy between theoretical predictions and experimental observations in LiPdH by performing carefully \emph{ab initio} calculations on the system, treating vibrational properties beyond the standard harmonic approximation. 
In Sec. \ref{sec_methods} we briefly overview the computational methods used in our work, while in Sec. \ref{sec_results} we present our results and explain the reason of the inconsistency between experimental and previous theoretical estimates. In Sec. \ref{sec_conclusions} we describe the main conclusions of the work.

\section{Computational Methods}

\label{sec_methods}

All calculations were performed using the plane-wave {\sc Quantum ESPRESSO} package \cite{Giannozzi2009, Giannozzi2017}, employing the Perdew-Zunger \cite{Perdew1981} parametrization of the exchange-correlation potential. The electron-ion interaction was represented using ultrasoft pseudopotentials with ${5s}^1$ and ${4d}^9$ electrons included in the valence of Pd.
A kinetic-energy cutoff of 120 Ry and 1000 Ry were set for the wave functions and the density, respectively. Brillouin zone integrations in the self-consistent calculations utilized a first-order Methfessel-Paxton \cite{Methfessel1989} smearing of 0.02 Ry broadening along with a 18 $\times$ 18 $\times$ 14 $\mathbf{k}$-point grid. Electronic properties, such as the electron localization function (ELF) \cite{Becke1990} and Bader charge \cite{bader1990} were computed using the postprocessing tools provided by {\sc Quantum ESPRESSO}.

The analysis of phonon behavior involved intricate computations. Initially, dynamical matrices were explicitly calculated on a dense 6 $\times$ 6 $\times$ 6 phonon $\mathbf{q}$-point grid using density functional perturbation theory (DFPT) \cite{Baroni1987} as implemented in {\sc Quantum ESPRESSO}. The calculations were performed with the same parameters described above. Anharmonic non-perturbative effects were subsequently included making use of the variational stochastic self-consistent harmonic approximation (SSCHA) method~\cite{Monacelli2021,Errea2013,Errea2014}. 
SSCHA calculations were performed at 300 K using a 2 $\times$ 2 $\times$ 2 supercell, and the dynamical matrices presented here are those obtained from the Hessian of the SSCHA free energy. To bridge the gap between dynamical matrices, which were initially obtained on a commensurate 2 $\times$ 2 $\times$ 2 $\mathbf{q}$-point grid, interpolation to a finer 6 $\times$ 6 $\times$ 6 grid was carried out. Specifically, this interpolation was performed by subtracting the harmonic dynamical matrices from the SSCHA dynamical matrices on the coarser grid, Fourier interpolating the resulting difference, and then adding this interpolated difference to the harmonic matrices on the finer grid. This approach ensures consistency between the harmonic and anharmonic representations at the desired resolution.

Finally, the electron-phonon interaction was scrutinized both at the harmonic and anharmonic levels. A 36 $\times$ 36 $\times$ 30 $\mathbf{k}$-point grid was employed for both cases, with a Gaussian smearing of 0.004 Ry to approximate the Dirac deltas. 
The critical temperature for superconductivity was determined by solving the Allen-Dynes modified McMillan equation~\cite{allen1975transition}, with ${\mu}^*= 0.1$.


\section{results}

\label{sec_results}

\subsection{LiPdH}

\begin{figure*}
\begin{center}
\includegraphics[width=2.0\columnwidth,draft=false]{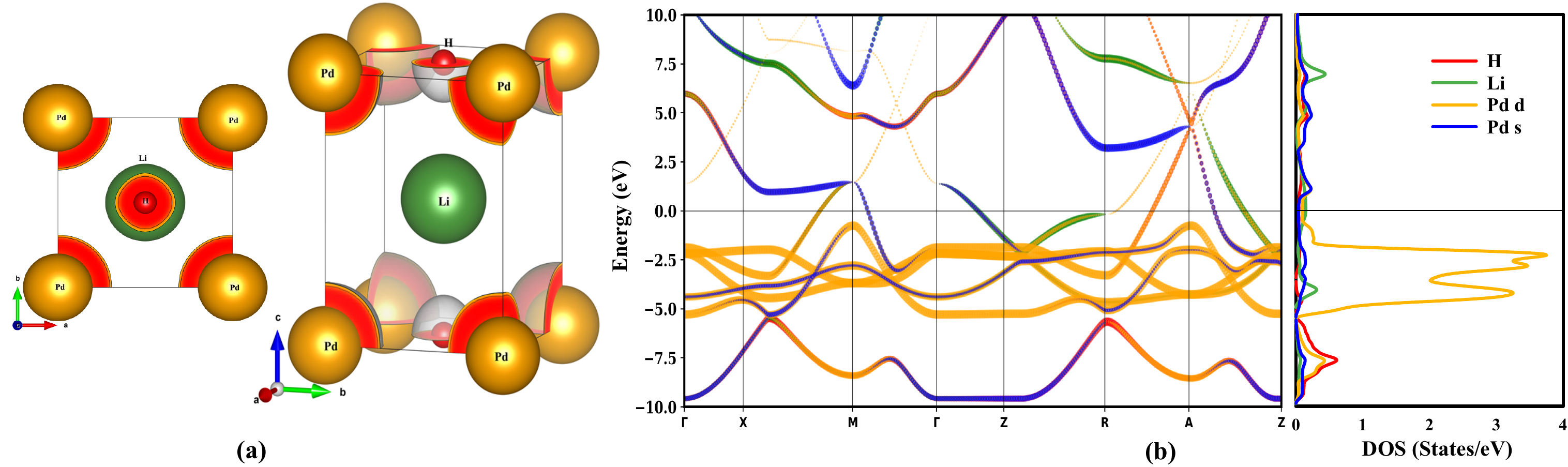}
\caption{(a) Crystal structure along the \emph c-axis (left) and the 3D structure (right), with the corresponding Electron Localization Function plotted on the structures at an isovalue of 0.8 for $P4/mmm$ LiPdH (Higher values of the ELF are depicted in red). (b) Electronic band structure (left panel) and density of states (DOS) (right panel) for $P4/mmm$ LiPdH. The projection onto atomic orbitals is also included.}

\label{Fig1}
\end{center}
\end{figure*}

The structural similarity of LiPdH to the cuprate family of superconductors was the starting point for studying its superconductivity\cite{singh1990possiblity}. The layered structure of LiPdH is predicted to be a superconductor due to its larger electron-phonon coupling constant compared to PdH. The tetragonal lattice of $P4/mmm$ LiPdH is shown in Fig. \ref{Fig1}(a), including
H and Pd atoms in layers that are separated by Li atoms. According to our density functional theory (DFT) relaxations at the Born-Oppenheimer level, the optimized lattice parameters have the values of $a=b= 2.75$ \r{A} and $c/a=1.399$. The Wyckoff positions of the atoms in the $P4/mmm$ structure are as follows: H occupies the $1a$ position (site symmetry $4/mmm$), Li occupies the $1b$ position (site symmetry $4/mmm$), and Pd occupies the $1c$ position (site symmetry $4/mmm$).

As shown in Fig.\ref{Fig1}(a), the crystal structure is depicted along the \emph c-axis on the left and as a 3D representation on the right. The corresponding ELF, calculated with an isovalue of 0.8, is visualized on both structures for the $P4/mmm$ LiPdH system, highlighting regions with higher ELF values in red. Li has a significant impact on the Pd-Pd bonding. In a comparable system such as PdH, the palladium atoms form an $fcc$ lattice, where they are metallically bonded at a distance of 2.03\r{A} \cite{long2018accounting}.
However, the inclusion of Li in PdH modifies this bonding pattern by introducing some Pd-H-Pd bridging bonds (top view of the structure along the \emph c-axis in Fig. \ref{Fig1} (a)), which affect the Pd-Pd bond lengths too. The Pd-Pd bonds, which are not mediated by hydrogen, remain nearly unchanged, while those that are part of the Pd-H-Pd bridges increase in length. According to our findings, the Pd-Pd distances are 2.70 \r{A}, whereas the Pd-Pd distances in the bridging Pd-H-Pd bonds are 2.85 \r{A}.

This modification in bonding has profound effects on the electronic structure and potentially the superconducting properties of  LiPdH. The band structure and the projected density of states (DOS) on atomic orbitals are depicted in Fig. \ref{Fig1}(b). LiPdH is a metal, where at the Fermi level Li and Pd states dominate the electronic density of states (DOS). The role of hydrogen at the Fermi level is around 10$\%$ of the total DOS, and it is more located in the energy range of 3 eV above the Fermi level. About 7 eV below the Fermi energy, the H 1\emph s mixes strongly with the Pd $d$ states, which is consistent with the bonding interactions in the PdH planes. Unlike PdH, in which the \textit{d} orbitals of palladium are located at the Fermi level \cite{long2018accounting}, in LiPdH this peak has been shifted to 2-5 eV below the Fermi level and the lithium atom predominantly contributes by sharing its valence electrons to form bonds at the Fermi level. The electron transfer mechanism is corroborated by a Bader charge analysis. This analysis indicates that in the $P4/mmm$ structure the Li atom loses approximately 0.83$e^-$ of charge, while the Pd and H atoms gain about 0.59$e^-$ and 0.23$e^-$, respectively, leading to ionic bonding characteristics.

\begin{figure*}
\begin{center}
\includegraphics[width=2.0\columnwidth,height=13cm, draft=false]{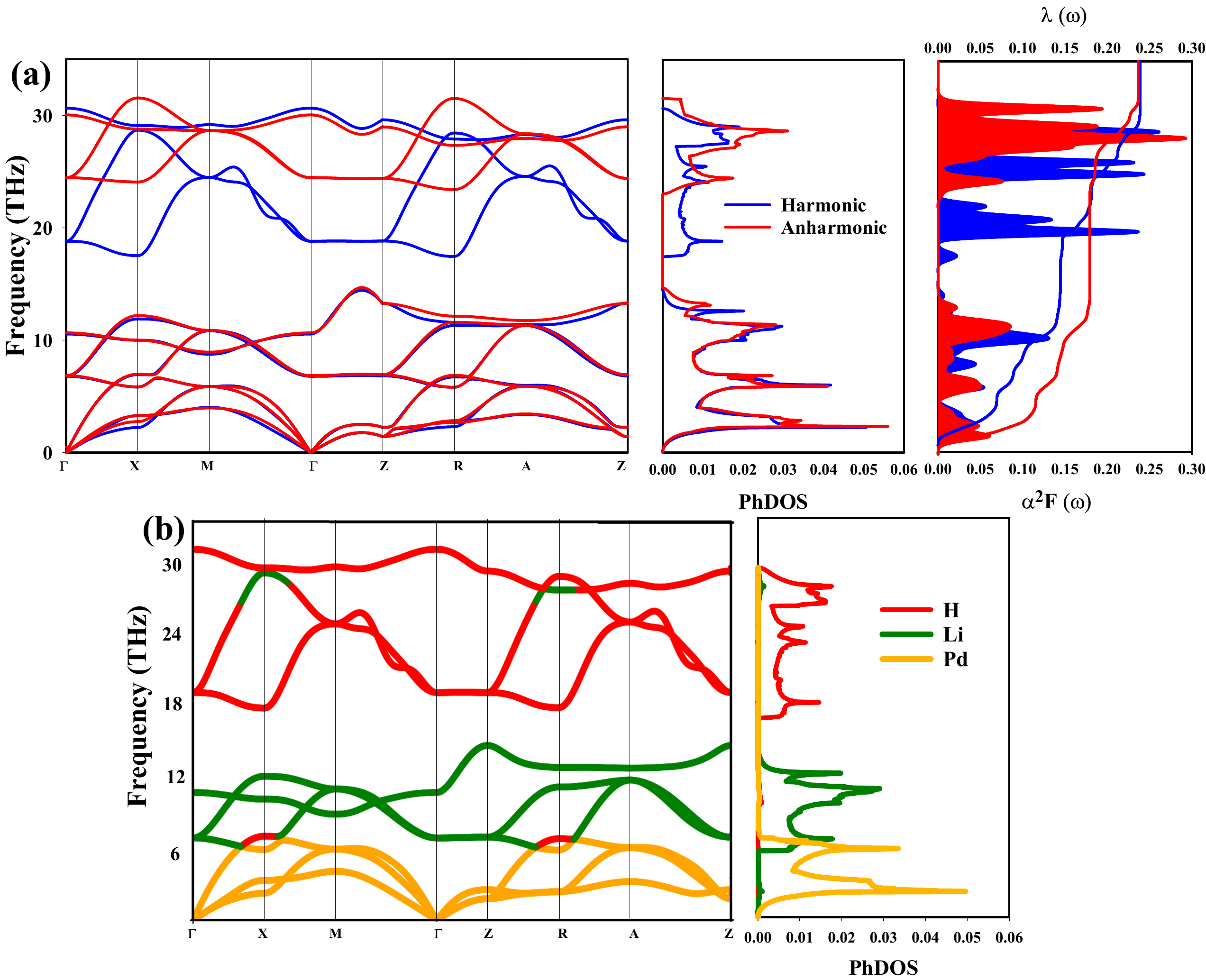}
\caption{(a) The phonon spectrum, the phonon density of states (PhDOS), Eliashberg function $\alpha^{2}F(\omega)$ and integrated electron-phonon coupling constants $\lambda(\omega)$ in the harmonic and anharmonic calculations for $P4/mmm$ LiPdH. (b) Projected phonon dispersions and DOS for $P4/mmm$ LiPdH structure obtained with the SSCHA at ambient pressure.}
\label{Fig2}
\end{center}
\end{figure*}

The phonon dispersion curves and phonon DOS (PhDOS) along high-symmetry directions of the Brillouin zone for LiPdH are shown in Fig. \ref{Fig2} (a), indicating the dynamical stability of LiPdH with no imaginary vibrational modes. This stability reflects the robustness of the structure at the harmonic level. The phonon spectrum reveals distinct acoustic and optical branches. The acoustic modes are primarily associated with the heavier Pd atoms, while the optical modes are dominated by Li and H vibrations, reflecting the significant mass disparity between the atoms in the structure.
At the harmonic level, the lower frequencies below 10 THz correspond to Li vibrations, while the H vibrations dominate the higher-frequency optical modes above 20 THz. Anharmonic corrections shift the hydrogen-related optical modes to higher frequencies and increase the separation between Pd- and Li-character modes from those dominated by hydrogen, while the low-energy modes in both approximations are similar. These corrections arise from the strong anharmonicity of hydrogen vibrations, which play a key role in the material’s vibrational dynamics.

Although hydrogen atoms in the layers are tightly bound in this structure, the amount of $x$ in LiPdH$_x$ significantly affects the \textit{c} parameter \cite{noreus1990absence}.
Consequently, the freedom of movement and weak coupling of lithium atoms with H-Pd planes result in larger displacements for lithium atoms than in-plane atoms. Consequently, the phonon frequencies of lithium are considerably softer than those of hydrogen, as shown in Fig. \ref{Fig2} (b). This is consistent with the ionic bonding of Li atoms, which is corroborated by the Eliashberg function $\alpha^2F(\omega)$, which does not show any particular enhancement around 10 THz. The Eliashberg function and the integrated $\lambda$ value indicate that the modes between 20-30 THz of hydrogen character have the most significant contribution to $\lambda$, as seen by comparing the PhDOS and the Eliashberg function $\alpha^2F(\omega)$. 


In PdH, the hydrogen modes are more concentrated in lower energy ranges (typically between 10-20 THz) \cite{Errea2013}. In contrast, in LiPdH, the hydrogen modes are primarily found in higher energy ranges (20-30 THz), where anharmonic effects tend to be less pronounced. As a result, the anharmonic renormalization in LiPdH is weaker compared to PdH, likely due to the higher frequency of the hydrogen modes in the system. This difference in the distribution of hydrogen modes and their interaction with anharmonicity is a key factor in the observed behavior of these materials. 

By integrating over the the Brillouin zone in both the harmonic and anharmonic approximations, we obtained $\lambda$ values of 0.24 and 0.23, respectively, both resulting in a transition temperature of less than 1 K.
The results in the harmonic and anharmonic approximations are presented for comparison with the calculated and measured results in Table \ref{tab1}.

Our \emph{ab initio} results are consistent with experimental works \cite{liu2018absence, noreus1990absence} and show notable differences from the previous computational estimates \cite{singh1990possiblity}. A crucial factor contributing to the inconsistency is related to the mean squared phonon frequencies used in the previous study. Since previous studies did not perform phonon calculations for LiPdH, and experimental and theoretical data were also unavailable, the values obtained for PdH were used instead of those of LiPdH. Additionally, as stated in \cite{singh1990possiblity} 
to estimate the T$_c$, the same values were applied for  H and Li, where $M_{\rm Li}\langle\omega^{2}\rangle_{\rm Li}=M_H\langle\omega^{2}\rangle_H$. Consequently, a large electron-phonon coupling was estimated, primarily driven by Li atoms, suggesting LiPdH as a promising candidate for superconductivity. Based on our calculations, we found that the values of $\langle\omega^2\rangle$ for Li, Pd, and H are 1.8, 5.36, and 1.24 $\text{eV}/\text{Å}^2$, respectively. These results lead to the corresponding values of $\lambda$, as shown in the fourth column of Table \ref{tab1}, according to the following relationship:
\begin{equation}
    \lambda_a=\frac{\eta_a}{M_a\langle\omega^2\rangle_a},
\end{equation}
where $\eta$ is the Hopfield parameter and $a$ is over different atoms, which is taken directly from Ref. \onlinecite{Park1995}. 
Notably, the calculated $\lambda$ values within McMillan-Hopfield's approach for Pd is very close to the one calculated fully from first principles here both in the harmonic and anharmonic calculations, and the value for H is within an acceptable range. However, for Li, there is a significant discrepancy between the value estimated here with the correct phonon average and the one estimated previously. The discrepancy in the T$_c$ observed in previous studies is thus primarily due to the oversimplification of the phonon frequency averages and the approximations made in the McMillan-Hopfield computational method, which in fact cearly overestimates $\lambda$. In the earlier work, the average phonon frequencies were estimated without fully accounting for the correct mass-dependent and atom-specific frequency contributions, leading to an overestimation of the transition temperature.



\begin{figure}
\hspace{-1cm}
\includegraphics[width=9.5cm, height=9cm]{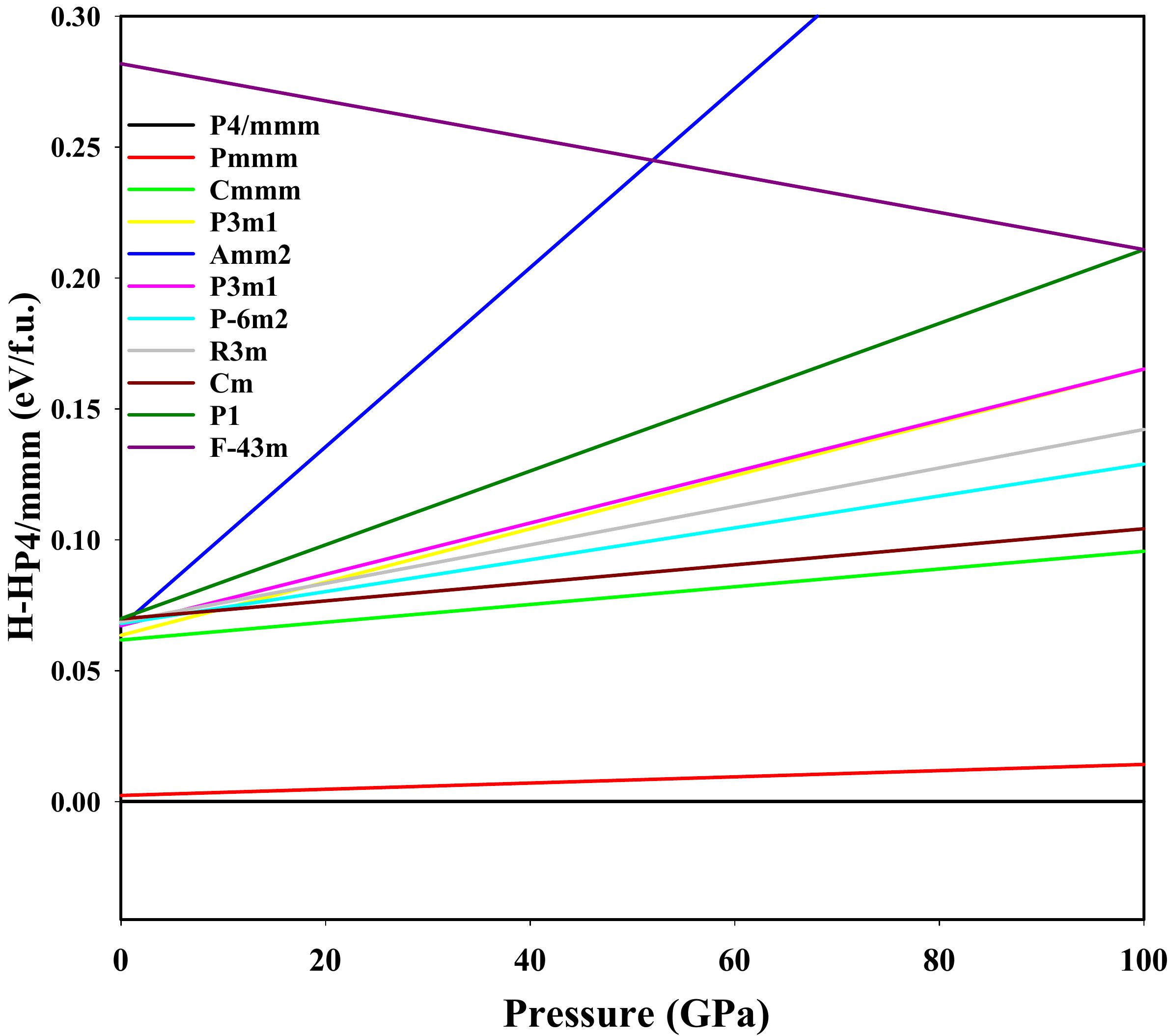}
\caption{Theoretical enthalpy versus pressure for LiPdH different structures obtained in the crystal structure prediction. The enthalpy of the $P4/mmm$ structure is the reference enthalpy.}
\label{Fig3}
\end{figure}

\subsection{Crystal structure prediction}


To predict the most stable structure and explore the possibility of new phases with potential high-T$_c$ superconductivity under pressure, we undertook a series of calculations aimed at understanding the material's behavior in response to varying pressure conditions. Given the complexity of pressure-induced phase transitions, it was important to systematically investigate the effects of pressure on the structural stability of the material. The main objectives of these simulations were to validate whether the $P4/mmm$ structure represents the ground state of LiPdH and to investigate whether the application of pressure could induce the formation of new phases that might exhibit superconducting properties at high temperatures. These simulations were particularly critical since high-pressure conditions often lead to unexpected phase transitions and novel material behaviors.
To investigate the low-lying crystal structures in the structure space, we performed total energy calculations in Quantum Espresso on a series of randomly generated structures obtained from CrySPY \cite{Yamashita2021}. This was followed by the relaxation of all internal degrees of freedom of the system. These simulations aim to elucidate the subtle structural changes occurring under high-pressure conditions, providing valuable insights into the complex behavior of the material.

\begin{table}
\begin{center}
 \begin{tabular}{|M{1cm}|M{1.7cm}|M{1.9cm}|M{1.7cm}|M{1cm}|M{1.3cm}|}
 \hline
 &This work (Harmonic) &This work (Anharmonic) & This work (within the McMillan-Hopleld equation) & Prev. Theo. Work\cite{singh1990possiblity} & Prev. Exp. Work\cite{noreus1990absence, Liu2017} \\ \hline
 \lambda$_H$ & 0.13 & 0.09 & 0.17 & 0.20 & - \\ 
 \lambda$_{Li}$ & 0.05 & 0.06 & 0.2 & 0.42 & - \\    
 \lambda$_{Pd}$ & 0.07 & 0.08 & 0.06 & 0.07 & - \\ 
 \lambda$_{Tot}$ & 0.24 & 0.23 & 0.43 & 0.7 & - \\      
 T$_c$ (K) & 0.1 & 0.06 & $<$10 & $>$10 & $<$2 \\ \hline
 \end{tabular} 
\caption{Calculated $\lambda$ and T$_c$ using the Allen-Dynes-modified McMillan formula with
the Coulomb pseudopotential parameter ${\mu}^*$ of 0.1 for comparison with the reported values for LiPdH.} 
\label{tab1}
\end{center}
\end{table}

Our crystal structure prediction calculations, summarized in Fig. \ref{Fig3}, focused on exploring the structural stability of the initial tetragonal phase up to 100 GPa. Remarkably, throughout this extensive pressure range, no phase transition was observed, and the $P4/mmm$ structure remained the dominant and stable configuration with the lowest enthalpy. This structure, as the initial configuration, displayed exceptional resilience and maintained its structural integrity without undergoing any phase transformations up to 100 GPa.

\begin{figure*}
\begin{center}
\includegraphics[width=2.0\columnwidth,height=6cm, draft=false]{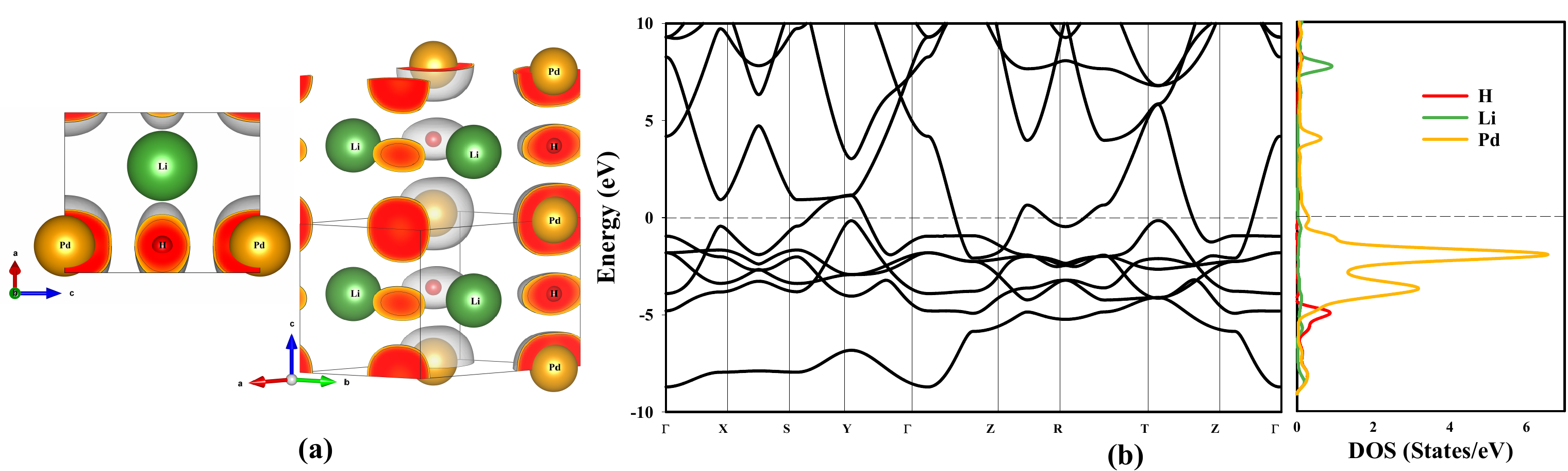}
\caption{(a) Crystal structure along the \emph b-axis (left) and the 3D structure (right), with the corresponding Electron Localization Function plotted on the structures at an isovalue of 0.8 for $Pmmm$ LiPdH (Higher values of the ELF are depicted in red). (b) Electronic band structure (left panel) and density of states (DOS) (right panel) for $Pmmm$ LiPdH.}
\label{Fig6}
\end{center}
\end{figure*}

While the $P4/mmm$ phase remains stable under high-pressure conditions (Fig. \ref{Fig3}), our simulations reveal the emergence of a $Pmmm$ structure with an enthalpy difference of only 2 meV per formula unit at 0 GPa compared to the $P4/mmm$ phase. Fig. \ref{Fig6} (a) illustrates the $Pmmm$ structure, where the Pd-H-Pd bond angles change dramatically from 
$90^\circ$ in the $P4/mmm$ to $180^\circ$ in $Pmmm$. This structural transformation significantly affects the electronic states near the Fermi level. Pd atoms play a more significant role in the Fermi surface and the density of states at the Fermi level decreases by approximately 3.2$\%$ compared to $P4/mmm$, indicating subtle modifications in the electronic behavior shown in Fig. \ref{Fig6} (b).

\begin{figure*}
\begin{center}
\includegraphics[width=2.0\columnwidth,height=7cm, draft=false]{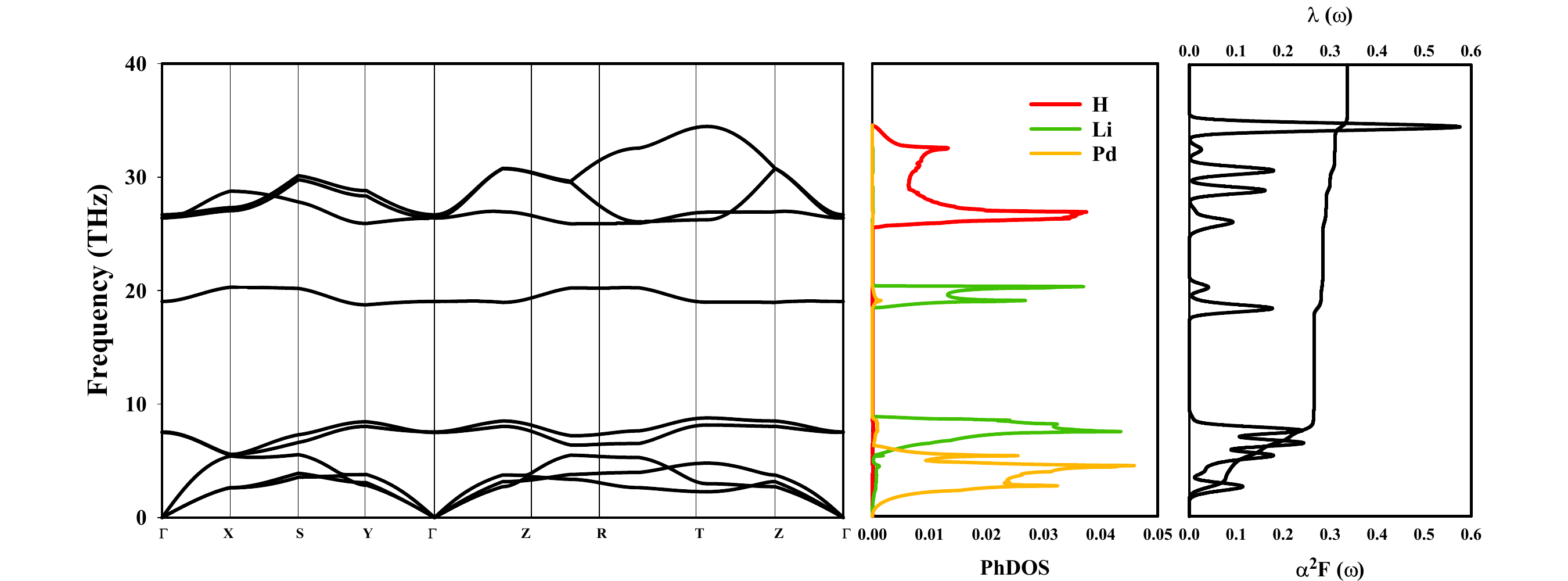}
\caption{The phonon spectrum, the phonon density of states (PhDOS), Eliashberg function $\alpha^{2}F(\omega)$, and integrated electron-phonon coupling constants $\lambda(\omega)$ in the harmonic calculations for $Pmmm$ LiPdH.}
\label{Fig7}
\end{center}
\end{figure*}

The dynamical stability of the $Pmmm$ phase is confirmed by its phonon spectrum shown in Fig. \ref{Fig7}. Hydrogen-related phonon modes exhibit noticeable softening compared to the $P4/mmm$ phase and this softening enhances the coupling between H vibrations and the Fermi-level electrons, as reflected in the Eliashberg spectral function. The electron-phonon coupling constant reaches 0.34, corresponding to a superconducting transition temperature of 0.8 K.

\subsection{$P4/mmm$ LiPdH at 50 and 100 GPa}

Our ab initio investigations revealed that the $P4/mmm$ LiPdH does not exhibit superconductivity under ambient pressure. This observation, however, does not discard that the electron-phonon coupling might be enhanced under pressure. Given that the $P4/mmm$ structure remains the most stable configuration up to 100 GPa as shown above, here we delve into a detailed analysis of its electronic and phonon properties at high pressures. 

\begin{figure*}
\begin{center}
\includegraphics[width=2.0\columnwidth,height=8cm, draft=false]{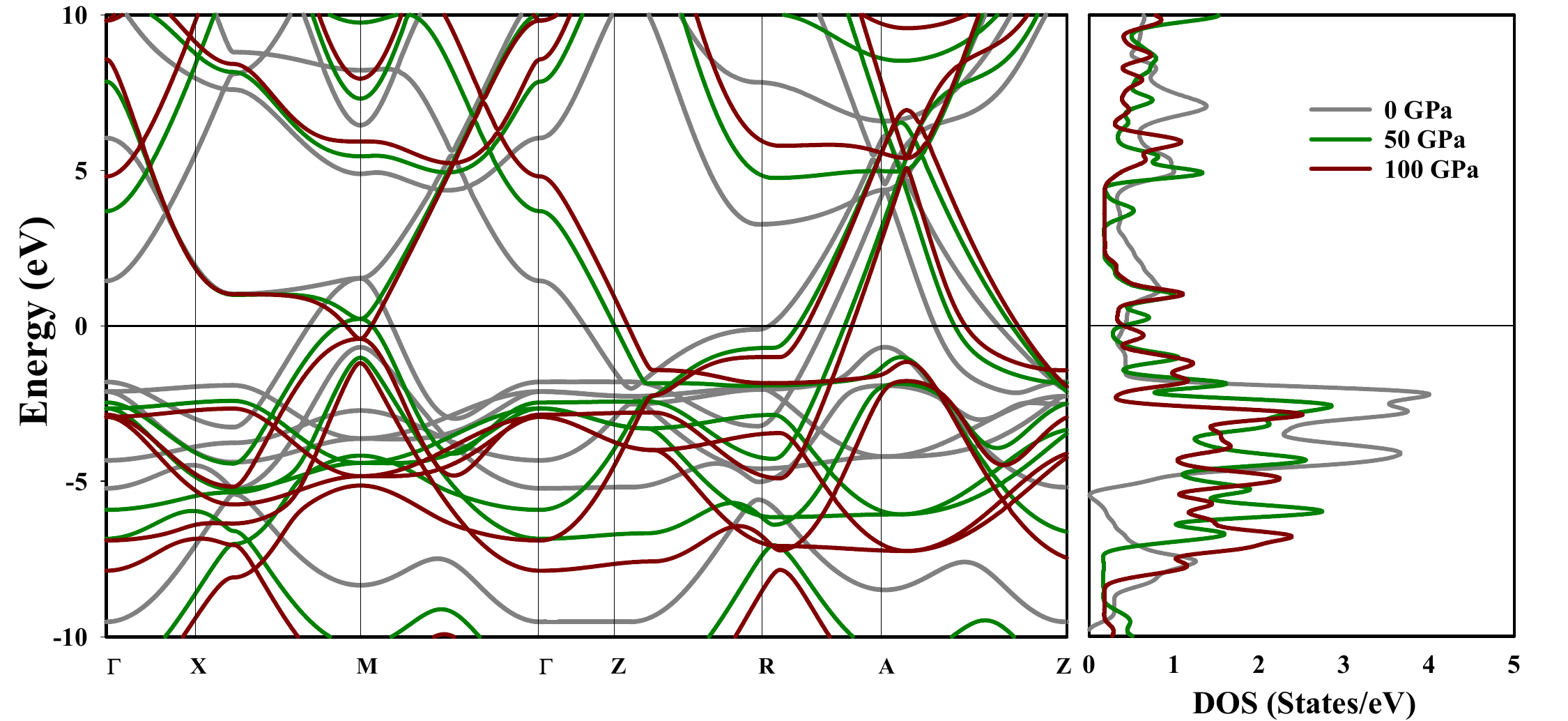}
\caption{ Electronic band structure (left panel) and density of states (DOS) (right panel) for $P4/mmm$ LiPdH at 0, 50, and 100 GPa.}
\label{Fig4}
\end{center}
\end{figure*}

The application of pressure induces significant changes in the interatomic distances. Reduction in volume typically leads to an increase in the overlap of electronic wave functions and contributes to higher phonon frequencies. The Li-H distance decreases by about 36$\%$, a reduction more substantial than the approximately 19$\%$ decrease observed between Pd and Li atoms.
This discrepancy arises primarily due to the smaller atomic radius of hydrogen. Consequently, the interactions between Li and H are strengthened under pressure, leading to alterations in their respective atomic charges. These modifications result in an increase of Bader charges for H, 0.26$e^-$ at 50 GPa and 0.27$e^-$ at 100 GPa. Specifically, the electron density around H atoms increased and reflected higher charge localization.
With increasing pressure, the Bader charge of hydrogen increases, indicating enhanced ionic characteristics in hydrogen, while the Bader charges of palladium and lithium decrease, reflecting a reduction in ionic properties in these elements. These changes, as shown in Table \ref{tab2}, suggest that the ionic interactions in hydrogen are strengthened under pressure. Therefore, we expect the phononic properties to decrease with increasing pressure, as ionic characteristics.
 \begin{table}
\begin{center}
 \begin{tabular}{|M{2cm}|M{1.5cm}|M{1.5cm}|M{1.5cm}|}
 \hline
 \diagbox[width=2cm,height=1cm]{Element}{Pressure} & 0 GPa & 50 GPa & 100 GPa  \\ \hline
 Pd & +0.59$e^-$ & +0.52$e^-$ & +0.48$e^-$ \\ 
 Li & -0.83$e^-$ & -0.78$e^-$ & -0.76$e^-$  \\    
 H & +0.23$e^-$ & +0.26$e^-$ & +0.27$e^-$  \\  \hline
 \end{tabular} 
\caption{Calculated Bader charge with increasing pressure.} 
\label{tab2}
\end{center}
\end{table}

The impact of pressure on superconductivity usually is nuanced and material-specific. In some cases, pressure can induce phase transitions, reduce bond lengths, enhance electronic interactions, and consequently enhance the electron-phonon coupling creating more favorable conditions for superconductivity. But in general, like in our case, the reverse can also occur. In Fig. \ref{Fig4}, as pressure is increased to 50 {GPa}, there is a significant decrease in the DOS at the Fermi level, amounting to a reduction of approximately 23$\%$. When pressure is further increased to 100 GPa, the density of states experiences an even more pronounced reduction, with a total decrease of about 47$\%$. 
 
\begin{figure*}
\begin{center}
\includegraphics[width=2.0\columnwidth,height=8cm, draft=false]{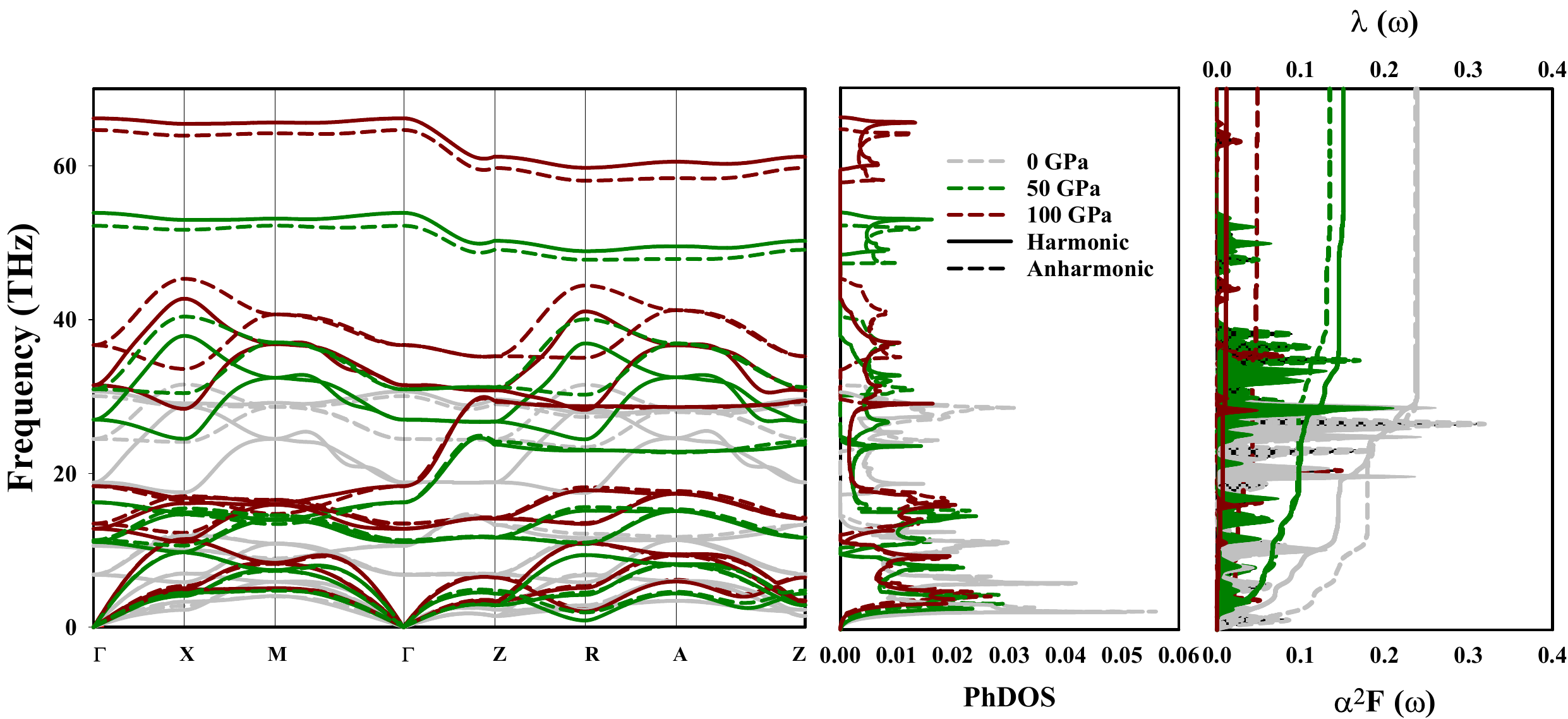}
\caption{The phonon spectrum, the phonon density of states (PhDOS), Eliashberg function $\alpha^{2}F(\omega)$ and integrated electron-phonon coupling constants $\lambda(\omega)$ for $P4/mmm$ LiPdH at 0, 50 and 100 GPa. Anharmonic phonons are used in this calculation and obtained with the SSCHA.}
\label{Fig5}
\end{center}
\end{figure*}

Generally, as expected, by applying pressure to a structure, the atoms in the lattice are forced closer together, the force constants are increased in the lattice (i.e., the forces that resist displacements of atoms from their equilibrium positions), and consequently, stiffer bonds shift frequencies to higher energies. In the case of $P4/mmm$ LiPdH, pressure decreases the density of states at the Fermi level and increases the phonon frequency, both of which influence the integrated $\lambda$. Referring to the anharmonic calculations in Fig. \ref{Fig5} (dashed lines), the soft phonon frequencies decrease with the application of pressure,  
but the anharmonic interaction is similar at all pressures, only affecting the highest-energy hydrogen-character optical modes. 
The reduction of superconductivity with increasing pressure is observed in the Eliashberg function, which shows broader peaks under pressure. As a result, the integrated electron-phonon coupling constant, $\lambda$, decreases significantly from 0.24 to 0.04 at 100 GPa. This change indicates the weakening of electron-phonon interactions due to the increased pressure, leading to a vanishing superconducting critical temperature.

\section{Conclusions}

\label{sec_conclusions}
 
Our ab initio calculations reveal that the prediction of superconductivity of the tetragonal LiPdH in the previous studies resulted from the oversimplified estimation of the average phonon frequencies, and uncovered the non-superconducting nature of this material observed experimentally. Additionally, by examining the enthalpy behavior under pressure, we concluded that the $P4/mmm$ remains the most stable structure up to 100 GPa. The application of pressure to this structure suppresses the density of states at the Fermi level and hardens phonon energies, which contribute to suppress the electron-phonon coupling constant.


 \section{ACKNOWLEDGMENTS}
We acknowledge valuable discussions with Đ. Dangić.  This work is based upon research funded by Iran National Science Foundation (INSF) under project No. 4003531.
I.E. acknowledges funding from ERC under the European Unions Horizon 2020 research and innovation program (Grant Agreements No. 802533); the Department of Education, Universities and Research of the Eusko Jaurlaritza, and the University of the Basque Country UPV/EHU (Grant No. IT1527-22); and the Spanish Ministerio de Ciencia e Innovación (Grant No. PID2022-142861NA-I00).


\end{document}